\documentclass[aps,prd,superscriptaddress,twocolumn,nofootinbib,10pt]{revtex4}
\usepackage[pdftex]{color}
\usepackage[dvipsnames]{xcolor}
\usepackage[sort&compress]{natbib}
\usepackage[colorlinks=true,linkcolor=RubineRed,filecolor=ForestGreen,urlcolor=ForestGreen,citecolor=ForestGreen,pdftex,plainpages=false]{hyperref}
\usepackage{url}
\pdfoutput=1
\usepackage{ifpdf}
\usepackage{color,hhline,psfrag,rotating}
\usepackage{dcolumn}
\usepackage{verbatim}
\usepackage{bm}
\usepackage{slashed}
\usepackage{graphicx}
\usepackage{amsfonts,amssymb,amsmath}
\usepackage{booktabs}  

\begin{document}
\title{QED interaction effects on heavy meson masses from lattice QCD+QED}

\author{D.~Hatton}
\email[]{daniel.hatton@glasgow.ac.uk}
\affiliation{SUPA, School of Physics and Astronomy, University of Glasgow, Glasgow, G12 8QQ, UK}
\author{C.~T.~H.~Davies}
\email[]{christine.davies@glasgow.ac.uk}
\affiliation{SUPA, School of Physics and Astronomy, University of Glasgow, Glasgow, G12 8QQ, UK}
\author{G.~P.~Lepage}
\affiliation{Laboratory for Elementary-Particle Physics, Cornell University, Ithaca, New York 14853, USA}
\collaboration{HPQCD collaboration}
\homepage{http://www.physics.gla.ac.uk/HPQCD}
\noaffiliation

\date{\today}

\begin{abstract}
Hadron masses are subject to few MeV corrections arising from QED interactions, almost entirely arising from the electric charge of the valence quarks. 
The QED effects include both self-energy contributions and 
interactions between the valence quarks/anti-quarks. 
By combining results from different signs of the valence quark electric charge 
we are able to isolate the interaction term which is 
dominated by the Coulomb piece, 
$\langle \alpha_{\mathrm{QED}}e_{q_1}e_{\overline{q}_2}/r \rangle$, 
in the nonrelativistic limit.
We study this for $D_s$, $\eta_c$ and $J/\psi$ mesons, working in lattice 
QCD plus quenched QED. We use gluon field configurations that include 
up, down, strange and charm quarks in the sea 
at multiple values of the lattice spacing. 
Our results, including also values for mesons with quarks heavier than charm, 
can be used to improve phenomenological models for 
the QED contributions. 
The QED interaction term carries
information about meson structure; we derive effective 
sizes $\langle 1/r_{\mathrm{eff}} \rangle^{-1}$ for 
$\eta_c$, $J/\psi$ and $D_s$ of 0.206(8) fm, 0.321(14) fm and 0.307(31) fm 
respectively. 
 
\end{abstract}

\maketitle
\section{Introduction}
\label{sec:intro}
Lattice QCD calculations can now achieve a very high level of accuracy for 
ground-state meson masses. For example, a recent calculation of the 
mass splitting between the $J/\psi$ and $\eta_c$ achieved an accuracy 
of 1 MeV~\cite{Hatton:2020qhk}. This precision requires that QED effects 
arising from the electric charge of the quarks be included in the 
calculation and this is now being widely 
done, with a variety of 
approaches~\cite{Borsanyi:2014jba, Giusti:2017dmp, Boyle:2017gzv, Basak:2018yzz, Kordov:2019oer, Hatton:2020qhk}. 

The QED effects arise almost entirely from the electric charge 
of the valence quarks. 
To $\mathcal{O}(\alpha_{\mathrm{QED}})$ 
we then expect the impact of QED on a meson made of quark $q_1$ and 
antiquark $\overline{q}_2$ to take the form 
\begin{equation}
\label{eq:QED-mass-shift}
\Delta M_{q_1\overline{q}_2} = Ae_{q_1}e_{\overline{q}_2} + Be_{q_2}^2 + Ce_{q_1}^2 ,
\end{equation} 
ignoring the much smaller effects from the electric charge of the 
sea quarks (suppressed by powers of $\alpha_s$ and sea quark mass
effects). 
Here $e_{q_1}$ and $e_{q_2}$ are the electric charges of quarks $q_1$ 
and $q_2$ in units of $e$, the magnitude of the charge on the electron. 
The last two terms are dominated by `self-energy' shifts 
in the valence quark masses. 
These are unphysical because they 
amount purely to a renormalisation of the quark mass by QED. 
The first term, with coefficient $A$, is physical, however. 
It is dominated, for nonrelativistic quarks, by the Coulomb interaction between 
the valence quark and antiquark in the meson. 
This effect depends on the average separation of the quarks 
and so provides a measure for the size of the meson. 
Its accurate determination requires a calculation 
that fully controls the QCD effects that bind the quark and antiquark 
into the meson, i.e. the use of lattice QCD.  

We will use lattice QCD calculations to which 
we also add the effect of QED on the valence quarks in 
an approach known as `quenched QED'~\cite{Duncan:1996xy}. 
This is simply achieved by generating a random photon field 
in momentum-space and then packaging the field in position space 
into a compact U(1) variable that can be multiplied into the gluon 
field as the Dirac equation is solved for each quark propagator. 
Since QCD is responsible for binding the quark and antiquark into 
the meson and the effect of QED is simply a perturbation to 
the meson mass, then 
the QED interaction term $Ae_{q_1}e_{\overline{q}_2}$ 
in Eq.~(\ref{eq:QED-mass-shift}) 
can be isolated by comparing 
results from lattice calculations in which we flip the sign 
of the electric charge for one of the quarks. 
We have
\begin{equation}
\label{eq:coulomb-isolate}
M(e_{q_1},e_{\overline{q}_2})-M(-e_{q_1},e_{\overline{q}_2}) = 2Ae_{q_1}e_{\overline{q}_2} .
\end{equation}

We focus here on studying this QED effect 
for relatively heavy mesons ($\eta_c$, $J/\psi$ and $D_s$) to 
test our understanding of the impact of QED. 
The reason for this is that 
the internal structure of these mesons is reasonably well understood and 
in the past we have made use of estimates of the Coulomb interaction 
effects to assess the impact of QED on these meson 
masses~\cite{Davies:2009tsa, Davies:2010ip}.  
In Section~\ref{sec:lattice} we describe our lattice calculation 
and the results and in Section~\ref{sec:discussion} we compare 
to these earlier estimates from phenomenological models. 
Section~\ref{sec:conclusions} gives 
our conclusions. 

\section{Lattice QCD calculation} \label{sec:lattice}

\begin{table*}
\centering
\caption{The parameters of the ensembles used in our calculation, 
numbered in column 1. Column 2 gives the QCD gauge coupling and 
column 3 the lattice spacing in units of the Wilson flow parameter, 
$w_0$~\cite{Borsanyi:2012zs}. The lattice spacing in fm is then given 
by using $w_0=$ 0.1715 fm, fixed from $f_{\pi}$~\cite{fkpi}. $L_s$ 
and $L_t$ are the lattice spatial and temporal extents in lattice units. 
Columns 6 and 7  give the sea light quark masses in lattice units, with 
the sea $u$ and $d$ quark masses taken to be the same and denoted $l$. 
Column 8 gives the valence $s$ quark mass used in the $D_s$ mesons. 
Columns 9 and 10 give the sea and valence $c$ quark masses in lattice 
units, respectively.  Not all sets are used for all calculations; * indicates 
that the set was used for charmonium, $\dag$ that the set was used for 
$D_s$ and $\ddag$ that the set was used for valence masses of $2m_c$. 
Column 11 gives the corresponding number of configurations used from the set. 
}
\label{tab:ensembles}
\begin{tabular}{lllllllllll}
\hline \hline
Set & $\beta$ & $w_0/a$ & $L_s$ & $L_t$ & $am_l^{\mathrm{sea}}$ & $am_s^{\mathrm{sea}}$ & $am_s^{\mathrm{val}}$ & $am_c^{\mathrm{sea}}$ & $am_c^{\mathrm{val}}$ & $N_{\mathrm{cfgs}}$ \\
\hline
1$*$ & 5.80 & 1.1272(7) & 24 & 48 & 0.0064 & 0.064 & - & 0.828 & 0.873 & 340 \\
2$\dagger$ & 5.80 & 1.1367(5) & 32 & 48 & 0.00235 & 0.0647 & 0.0677 & 0.831 & 0.863 & 100 \\
3$*\dagger$ & 6.00 & 1.4029(9) & 32 & 64 & 0.00507 & 0.0507 & 0.0533 & 0.628 & 0.650 & 220$*$ / 140$\dagger$ \\
4$*$ & 6.30 & 1.9330(20) & 48 & 96 & 0.00363 & 0.0363 & - & 0.430 & 0.439 & 371 \\
5$\dagger \ddagger$ & 6.30 & 1.9518(7) & 64 & 96 & 0.00120 & 0.0363 & 0.036 & 0.432 & 0.433 & 87$\dagger$ / 184$\ddagger$ \\
6$*\dagger \ddagger$ & 6.72 & 2.8960(60) & 48 & 144 & 0.0048 & 0.024 & 0.0234 & 0.286 & 0.274 & 133$*$ / 87$\dagger$ / 199$\ddagger$ \\
\hline \hline
\end{tabular}
\end{table*}

\begin{table}
\centering
\caption{Results for the charmonium case. 
Column 2 gives the ground-state $\eta_c$ (upper rows) and $J/\psi$ (lower rows) 
meson masses in lattice units in the pure QCD case for the gluon 
field configuration sets given in column 1. Column 3 gives the ratio 
of the mass difference for the physical and unphysical QED scenarios 
(see Eq.~(\ref{eq:QCDQEDrat})) to the pure QCD mass. Column 4 gives the finite-volume 
correction needed on that gluon configuration set for the unphysical 
QED scenario (Eq.~(\ref{eq:fvolshift}) for meson charge 4$e$/3). 
The uncertainty in $\Delta_{\mathrm{FV}}$ comes mainly from the 
uncertainty in the lattice spacing and does not include the systematic 
error from missing higher orders in $1/L_s$ (see text). 
Finally column 5 gives the extracted coefficient, $A_{\eta_c}$ 
or $A_{J/\psi}$ (Eq.~(\ref{eq:calcA})). 
}
\label{tab:charm-results}
\begin{tabular}{lllll}
\hline \hline
Set & $aM^{\text{QCD}}_{\eta_c}$ & $R_{\eta_c}$ & $\Delta_{\mathrm{FV}}$ [MeV] & $A_{\eta_c}$ [MeV] \\
\hline
1 & 2.305364(39) & -0.002080(39) & -1.0308(54) & 8.16(14) \\
3 & 1.848041(35) & -0.001806(25) & -0.9600(51) & 7.139(91) \\
4 & 1.342455(21) & -0.0017726(58) & -0.8795(47) & 6.944(42) \\
6 & 0.896675(24) & -0.001641(21) & -1.3373(75) & 7.020(80) \\
\hline
Set & $aM^{\text{QCD}}_{J/\psi}$ & $R_{J/\psi}$ & $\Delta_{\mathrm{FV}}$ [MeV] & $A_{J/\psi}$ [MeV] \\
\hline
1 & 2.39308(14) & -0.001342(23) & -1.0295(54) & 5.844(85) \\
3 & 1.914749(67) & -0.001144(12) & -0.9589(51) & 5.057(77) \\
4 & 1.391390(43) & -0.001063(17) & -0.8785(47) & 4.688(63) \\
6 & 0.929860(54) & -0.000883(25) & -1.3352(75) & 4.580(90) \\
\hline \hline
\end{tabular}
\end{table}

\begin{table}
\centering
\caption{Results, as in Table~\ref{tab:charm-results}, 
but now for heavyonium mesons using 
quarks with mass $2m_c$. 
Column 2 gives the ground-state `$\eta_{2c}$' (upper rows) and `$\psi_{2c}$' (lower rows) 
meson masses in lattice units in the pure QCD case for the gluon 
field configuration sets given in column 1. Column 3 gives the ratio 
of the mass difference for the physical and unphysical QED scenarios 
to the pure QCD mass. Column 4 gives the finite-volume 
correction needed on that gluon configuration set for the unphysical 
QED (Q=$4e/3$) scenario. Finally column 5 gives the extracted coefficient, $A_{\eta_{2c}}$ 
or $A_{\psi_{2c}}$. 
}
\label{tab:heavy-results}
\begin{tabular}{lllll}
\hline \hline
Set & $aM^{\text{QCD}}_{\eta_{2c}}$ & $R_{\eta_{2c}}$ & $\Delta_{\mathrm{FV}}$ [MeV] & $A_{\eta_{2c}}$ [MeV] \\
\hline
5$\ddagger$ & 2.185464(53) & -0.001527(22) & -0.6552(34) & 9.17(13) \\
6$\ddagger$ & 1.487111(36) & -0.0012878(80) & -1.3137(74) & 8.657(66) \\
\hline
Set & $aM^{\text{QCD}}_{\psi_{2c}}$ & $R_{\psi_{2c}}$ & $\Delta_{\mathrm{FV}}$ [MeV] & $A_{\psi_{2c}}$ [MeV] \\
\hline
5$\ddagger$ & 2.221922(47) & -0.001078(12) & -0.6550(34) & 6.789(76) \\
6$\ddagger$ & 1.509707(53) & -0.000886(11) & -1.3132(74) & 6.491(72) \\
\hline \hline
\end{tabular}
\end{table}

We work on $n_f=2+1+1$ gluon field configurations generated 
by the MILC collaboration~\cite{Bazavov:2012xda,Bazavov:2017lyh}. 
These configurations include the effect of $u/d$, $s$ and $c$ 
quarks in the sea using the Highly Improved Staggered Quark (HISQ) 
action~\cite{Follana:2006rc}. Details of the parameters for the configurations are 
given in Table~\ref{tab:ensembles}. 

In~\cite{Hatton:2020qhk} we 
analysed charmonium correlators calculated in pure QCD and 
in QCD + quenched QED using these (and further sets) of 
gluon field configurations. This enabled us to determine 
accurately how the $\eta_c$ and $J/\psi$ meson masses shift 
(for a fixed valence $c$ quark mass) 
when the $2e/3$ electric charge of the valence $c$ quarks is 
included.  The shifts are very small, upwards by $\sim$0.1\%, but clearly 
visible. From this we could work out how the $c$ quark mass 
should be retuned when QED is switched on. We chose the 
natural tuning procedure in which the $c$ quark mass 
is adjusted in both QCD and QCD+QED until the $J/\psi$ meson mass 
determined on the lattice agrees with experiment. 
This led us to a determination of the $c$ quark mass in 
the $\overline{\text{MS}}$ scheme of 
$\overline{m}_c(3\,\text{GeV})_{\text{QCD+QED}} =$0.9841(51) GeV. 
This value is then 0.2\% lower than in pure QCD~\cite{Hatton:2020qhk}. 

The reason that the inclusion of QED lowers the $c$ quark mass 
(tuning to a fixed meson mass) is because the positive self-energy terms  
in Eq.~(\ref{eq:QED-mass-shift}) raise the meson mass. The Coulomb interaction  
is attractive inside a charmonium meson, however, and so must lower 
the meson mass. Here we set out to isolate the Coulomb-dominated 
piece of the QED 
effect. 

As described in Section~\ref{sec:intro} we can do this by comparing 
two calculations in QCD+QED (which will be shorthand for QCD + quenched QED in what 
follows). One calculation is the normal QCD+QED charmonium calculation with 
$c$ and $\overline{c}$ quarks with opposite electric charge. 
The second calculation is one in which the $c$ quark electric charge is 
flipped but not that of the $\overline{c}$. The difference between the two 
results then gives twice the QED interaction contribution to the meson 
mass (Eq.~(\ref{eq:coulomb-isolate})). 

Note that the second calculation is for an unphysical scenario 
as far as QED is concerned. The underlying QCD physics is the 
same in both cases. We use the same valence quark mass in 
the two calculations, i.e. a mass close to the tuned $c$ quark mass    
in QCD+QED for the physical scenario. 
Our valence $c$ quark masses are given in Table~\ref{tab:ensembles}. These are the same 
as masses used in~\cite{Hatton:2020qhk}, with tuning errors below 0.5\%. 

For the two calculations we combine $c$ and $\overline{c}$ propagators 
to generate two-point correlation functions that we average over 
the gluon field configurations. We fit these as a function of time 
separation between source and sink to determine the ground-state masses 
in lattice units. The procedure for including quenched QED and 
for fitting correlation functions is exactly the same as that 
described in~\cite{Hatton:2020qhk} and we do not repeat the
discussion of either procedure here.  

In Table~\ref{tab:charm-results} we give our results for the
$\eta_c$ and $J/\psi$ mesons. We calculate the ground-state masses 
in the pure QCD case and also in the physical and unphysical QCD + 
quenched QED scenarios. It is convenient to give the QCD+QED results 
for the masses as a ratio to the value in pure QCD. 
In Table~\ref{tab:charm-results} we 
therefore give values for the difference of the ratios in the physical 
and unphysical QCD+QED cases:
\begin{equation}
\label{eq:QCDQEDrat} 
R = \frac{M(e_{q_1},e_{\overline{q}_2})-M(-e_{q_1},e_{\overline{q}_2})}{M(0,0)}.
\end{equation}
$e_{q_1}$ and $e_{\overline{q}_2}$ are the electric charges of 
the quark and antiquark in units of $e$; for the 
charmonium case these are $2/3$ and $-2/3$. 
Multiplying $R$ by the mass in the pure QCD case, $M(0,0)$ (column 2 of 
Table~\ref{tab:charm-results}), then gives 
the mass difference needed for Eq.~(\ref{eq:coulomb-isolate}).

\begin{figure}
\centering
\includegraphics[width=0.45\textwidth]{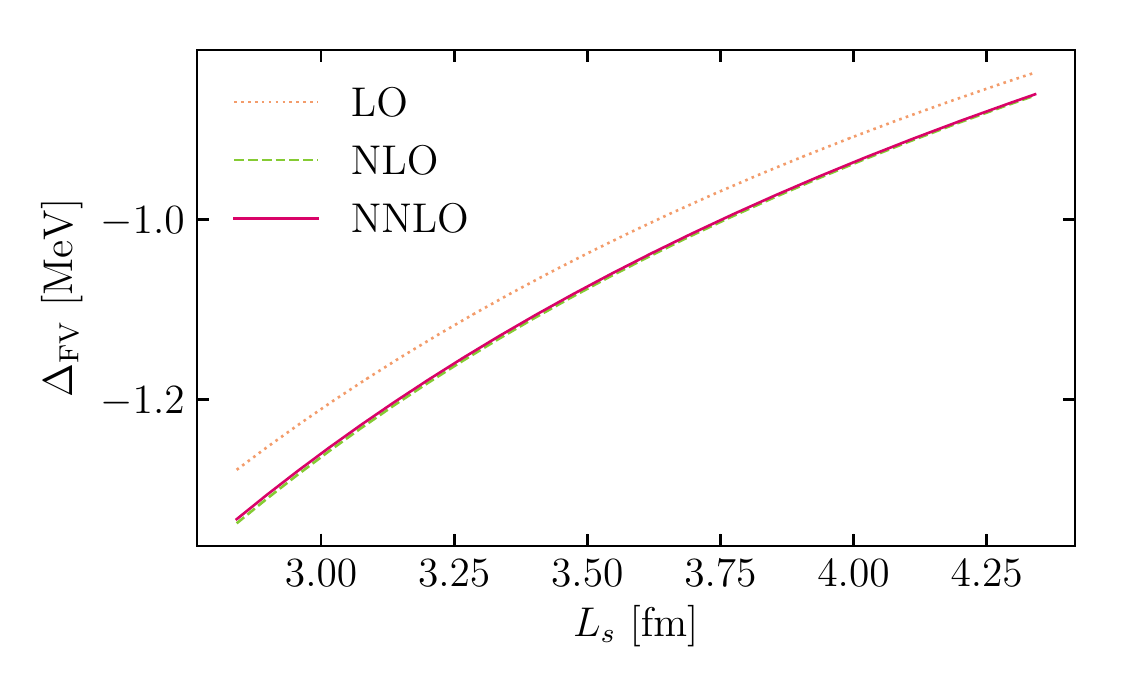}
\caption{A plot to show the size of finite-volume shifts needed 
in the $\eta_c$ case (for the unphysical QED scenario with 
meson charge $Q=4/3$) as a function of lattice spatial size. 
The plot compares the leading-order $1/L_s$ calculation, which 
is independent of meson mass, to the 
result of adding in higher order terms in $1/L_s$.  
}
\label{fig:fv}
\end{figure}

\begin{figure}
\centering
\includegraphics[width=0.45\textwidth]{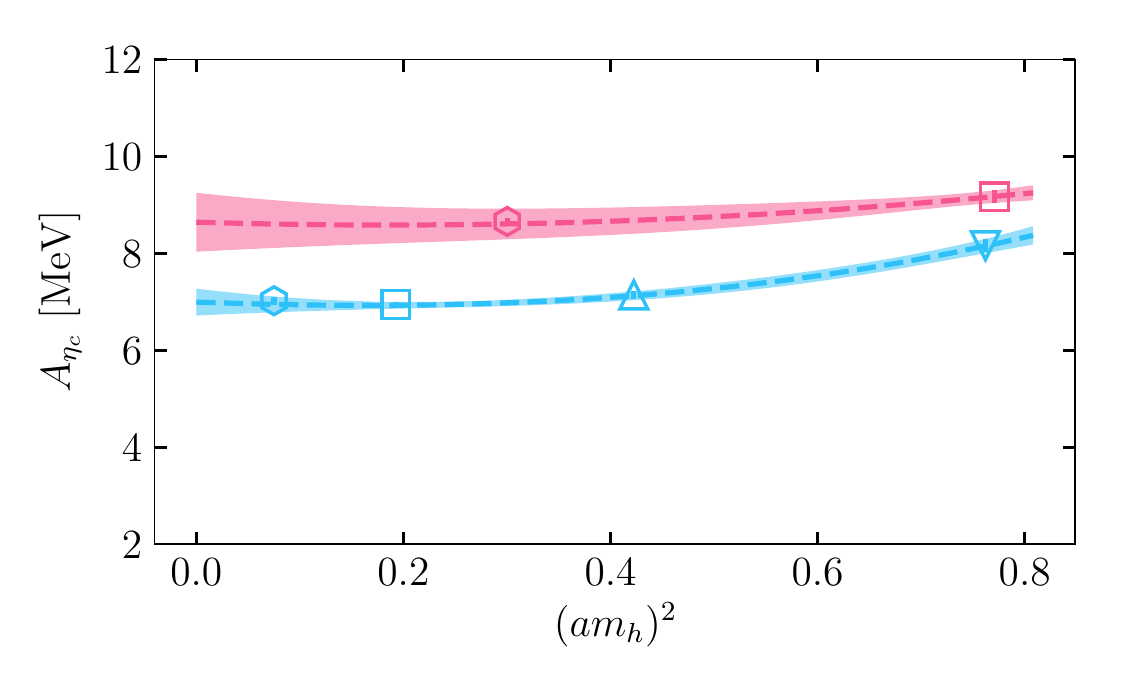}
\includegraphics[width=0.45\textwidth]{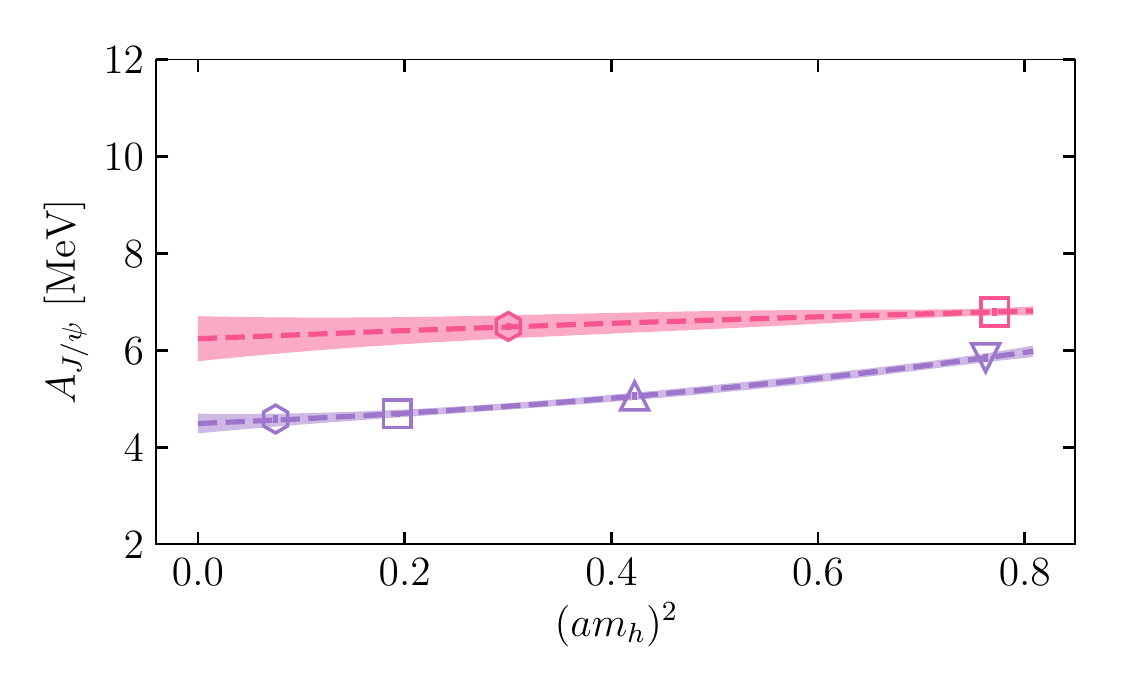}
\caption{The coefficient, $A$, of the QED interaction effect 
(Eq.~(\ref{eq:QED-mass-shift})) in the 
$\eta_c$ (upper plot, blue symbols) and $J/\psi$ 
(lower plot, purple symbols) meson masses, shown 
as a function of squared lattice spacing (given in units of the quark 
mass, denoted $m_h$ but here $m_c$).  
The fit is described in the text and shown by the curves in each plot. 
The pink symbols show the same 
results, but for mesons made from a quark-antiquark pair with quark 
mass $m_h$ twice that of the $c$ quark. The symbol shape denotes the gluon 
field configurations used and is the same as that for the matching 
charmonium calculation.  
}
\label{fig:charmonium}
\end{figure}

One difference between the physical and unphysical QED scenarios that we must take 
into account, however, is that of finite-volume effects from QED. In the physical 
scenario the charmonium meson is electrically neutral and finite-volume 
effects are negligible, as demonstrated in~\cite{Hatton:2020qhk}. 
In the unphysical scenario the meson has an electric charge of $4e/3$ and 
QED finite-volume effects are much larger. The finite-volume effects 
have been calculated analytically as an inverse power series in the spatial 
extent of the lattice~\cite{Hayakawa:2008an, Davoudi:2014qua, Borsanyi:2014jba}. 
We need only the (universal) result up to $1/L_s^2$ which takes the form 
\begin{eqnarray}
\label{eq:fvolshift}
\Delta_{\mathrm{FV}}(L_s) &=& M(L_s) - M(\infty) \\
&=& -\frac{Q^2\alpha_{\text{QED}}\kappa}{2L_s}\left( 1+ \frac{2}{M L_s}\right) \nonumber
\end{eqnarray}
with $\kappa =$ 2.8373 and $Q$ the meson electric charge in units of $e$. 
The leading term, which is independent of meson mass, 
takes a value of $Q^2 \times$ 0.5 MeV on a 4 fm lattice. We see then that the 
finite-volume effects are small here, but not negligible compared 
to our QED shifts. We handle them by correcting our 
finite-volume masses using the formula above for the cases where we have an 
(unphysical) electrically charged charmonium meson.  
The finite-volume shifts in each case are given in Table~\ref{tab:charm-results}.

Figure~\ref{fig:fv} plots $\Delta_{\mathrm{FV}}$ for $Q=4/3$ as a function of 
spatial lattice size for the range of lattice sizes that we use here. 
The plot compares the leading $1/L_s$ term of Eq.~(\ref{eq:fvolshift}) 
to the result of including both the $1/L_s$ and $1/L_s^2$ terms. 
We also show the impact of next-to-next-to-leading-order (NNLO) terms 
at $1/L_s^3$ from~\cite{Davoudi:2014qua}. We take the value of 
$\langle r^2 \rangle$ that appears in the $1/L_s^3$ terms from 
vector meson dominance as $6/M_{J/\psi}^2$ 
(since we have shown in~\cite{Davies:2019nut} that vector dominance works well 
for the electromagnetic form factor of mesons at small 
momentum-transfer, including for the $\eta_c$).  
We estimate the systematic 
uncertainty from missing out the $1/L_s^3$ terms at 0.005 MeV, which 
is negligible compared to other sources of uncertainty. 

We then combine the mass differences and finite-volume shifts 
to isolate the QED interaction effect for the $\eta_c$ and 
$J/\psi$ (Eq.~\ref{eq:coulomb-isolate}). 
The coefficient, $A$, is determined as: 
\begin{equation}
\label{eq:calcA}
A = \frac{1}{2e_{q_1}e_{\overline{q}_2}} \left(R \times M(0,0) + \Delta_{\mathrm{FV}}\right). 
\end{equation}
These values are given for each ensemble in Table~\ref{tab:charm-results}. 

We plot $A_{\eta_c}$ and $A_{J/\psi}$ in Fig.~\ref{fig:charmonium} 
as a function of lattice spacing. 
We see that, as expected, the attractive Coulomb interaction yields 
a negative contribution to the meson masses because $A$ is positive and 
$e_{q_1}e_{\overline{q}_2}$ is negative. The $A$ values are 
not the same for the $\eta_c$ and $J/\psi$ mesons because of the 
QED hyperfine interaction, which acts in the same direction as the 
QCD hyperfine interaction raising the $J/\psi$ mass relative to the 
$\eta_c$~\cite{Hatton:2020qhk}. 

In order to obtain a value for the coefficient $A$ in the continuum limit we use 
a fit that allows for discretisation errors as well as possible effects from 
the mistuning of the charm quark valence mass and the mistunings of the sea 
quark masses from their physical values. 
The fit form we use is similar to that in \cite{Hatton:2020qhk}:
\begin{eqnarray}
\label{eq:amc-fit}
A(a^2, \delta m) &=& A \Bigg[ 1 + \sum_{i=1}^3 c_a^{(i)} (am_c)^{2i} + c_{m,\mathrm{sea}} \delta_m^{\mathrm{sea},uds} \nonumber \\ 
&+& c_{c,\mathrm{sea}} \delta_m^{\mathrm{sea},c} + c_{c,\mathrm{val}} \delta_m^{\mathrm{val},c} \Bigg] . 
\end{eqnarray}
The mass mistuning terms here are defined as in \cite{Hatton:2020qhk}:
\begin{eqnarray}
\label{eq:mistunings}
\delta_m^{\mathrm{sea},uds} &=& \frac{2m_l^{\mathrm{sea}} + m_s^{\mathrm{sea}} - 2m_l^{\mathrm{phys}} - m_s^{\mathrm{phys}}}{10m_s^{\mathrm{phys}}} , \\
\delta_m^{\mathrm{sea},c} &=& \frac{m_c^{\mathrm{sea}} - m_c^{\mathrm{phys}}}{m_c^{\mathrm{phys}}} , \nonumber \\
\delta_m^{\mathrm{val},c} &=& \frac{M_{J/\psi} - M_{J/\psi}^{\mathrm{expt}}}{M_{J/\psi}^{\mathrm{expt}}} . \nonumber
\end{eqnarray}
$M_{J/\psi}$ is the lattice value in the QCD+QED case with the physical QED scenario. 
For the experimental $J/\psi$ mass we use 3.0969 GeV \cite{Tanabashi:2018oca}.
We use priors of 0(1) for the $c_a^{(i)}$, $c_{m,\mathrm{sea}}$ and 
$c_{c,\mathrm{val}}$ coefficients and a prior of $0\pm 0.1$ for $c_{c,\mathrm{sea}}$.
The mistuning terms in the fit have very little effect but including 
them allows us to incorporate uncertainties from them in the final result.

With $\chi^2/\mathrm{dof}$  of 0.06 and 0.1 respectively we find 
\begin{eqnarray}
\label{eq:charmres}
A_{\eta_c} &=&  6.99(28) \, \mathrm{MeV} \\
A_{J/\psi} &=&  4.49(20) \, \mathrm{MeV} \, .\nonumber 
\end{eqnarray}
The uncertainty is dominated by that from the extrapolation to 
zero lattice spacing and is much larger than that from possible 
systematic errors in the finite-volume correction discussed above. 
The fit is able to pin down the coefficient of the $(am_c)^2$ term 
($c_a^{(1)}$) to be within 0.3 of zero. This is consistent with 
the expectation that this coefficient should be of 
size $\mathcal{O}(\alpha_s)$~\cite{Follana:2006rc}. 

The Coulomb interaction effect probes the internal structure of the 
meson at short distances between the quark-antiquark pair. It is therefore 
interesting to ask how the coefficient $A$ changes for heavier 
quarks than the $c$ quark. In Table~\ref{tab:heavy-results} we give our results for a 
heavyonium meson made from a quark-antiquark pair with quark mass 
twice that of the $c$ quark (but the same electric charge). 
Again we use these results to determine the coefficient $A$ (which is 
independent of electric charge) in this case. These results are 
also plotted in Fig.~\ref{fig:charmonium}. The coefficient $A$ 
is substantially larger for the heavier mass case.   

We perform fits to the heavier mass points 
also using the fit form of Eq.~\eqref{eq:amc-fit}, but with $am_c$ now 
replaced with $2am_c$ and dropping the $a^6$ terms because we have 
results on fewer ensembles for this case. 
The functional form of the lattice spacing dependence 
should be the same in the $m_c$ and $2m_c$ cases up to possible 
dependence on the squared velocity of the heavy quark inside 
the bound-state in higher order coefficients in $a^2$~\cite{Follana:2006rc}. 
We therefore use the results of the $m_c$ fit as prior information 
to constrain the coefficients 
of the lattice spacing dependence ($c_a^{(1)}$ and $c_a^{(2)}$) 
in the fit for the $2m_c$ case. 
This amounts to choosing a prior width of 0.3 for the $c_a^{(1)}$ coefficient 
and 0.7 for $c_a^{(2)}$. 
We find 
\begin{eqnarray}
\label{eq:2charmres}
A_{\eta_{2c}} &=&  8.64(61) \, \mathrm{MeV} \\
A_{\psi_{2c}} &=&  6.24(27) \, \mathrm{MeV} . \nonumber 
\end{eqnarray}
The fits give $\chi^2/\mathrm{dof}$ of 0.39 and 0.09 respectively.

We will discuss a comparison of the 
values for $A$ for charmonium and heavyonium 
with those determined from static QCD potentials 
in Section~\ref{sec:discussion}.  

We can contrast the heavyonium case with that of 
a heavy-light meson. The simplest meson to use for this case 
is the heavy-strange meson since this has no valence light quark. 
We carry out the same analysis for the $D_s$ as for charmonium, 
but now the QED finite-volume 
effects (Eq.~(\ref{eq:fvolshift})) apply to both the physical 
scenario (since the $D_s$ meson is electrically charged with $Q$=1) and 
the unphysical scenario (where the `$D_s$' has the smaller charge $Q=1/3$).  
In Eq.~(\ref{eq:calcA}) we therefore substitute for $\Delta_{\mathrm{FV}}$ 
the difference $\delta \Delta_{\mathrm{FV}}$ of the finite-volume effects 
for the $Q=1$ and $Q=1/3$ cases. 
The valence $s$ quark masses that we use are given in Table~\ref{tab:ensembles} 
and are those obtained from the $m_s$ tuning exercise in~\cite{Chakraborty:2014aca}. 
Our results for the $D_s$ meson mass in pure QCD, along with the ratio 
$R$ of Eq.~(\ref{eq:QCDQEDrat}) and the finite-volume shifts discussed above, 
are given in Table~\ref{tab:dsresults}. 

\begin{table}
\centering
\caption{ Results that we use to obtain the QED
interaction effect for the $D_s$ meson. 
Column 2 gives the ground-state $D_s$ 
meson mass in lattice units in the pure QCD case for the gluon 
field configuration sets given in column 1. Column 3 gives the ratio 
of the mass difference for the physical and unphysical QED scenarios 
(Eq.~(\ref{eq:QCDQEDrat})) to the pure QCD mass. Column 4 gives the finite-volume 
correction needed on that gluon configuration set for the difference 
between the physical QED scenario (with meson charge 1) and the unphysical 
QED scenario (with meson charge 1/3). Finally column 5 gives the 
extracted coefficient of the effect on the mass from the quark electric 
charge interaction term, $A_{D_s}$. 
}
\label{tab:dsresults}
\begin{tabular}{lllll}
\hline \hline
Set & $aM^{\text{QCD}}_{D_s}$ & $R_{D_s}$ & $\delta \Delta_{\mathrm{FV}}$ [MeV]  & $A_{D_s}$ [MeV] \\
\hline
2 & 1.52428(16) & 0.000921(39) &  0.3917(24) & 5.01(18) \\
3 & 1.22386(17) & 0.000820(29) &  0.4880(30) & 4.74(13) \\
5 & 0.87740(10) & 0.000891(47) &  0.3345(20) & 4.70(21) \\
6 & 0.59203(22) & 0.00075(12) &  0.6839(46) & 4.86(52) \\
\hline \hline
\end{tabular}
\end{table}

\begin{table}
\centering
\caption{Results as for Table~\ref{tab:dsresults} but now for a heavy quark mass
with value $2m_c$, for which we denote the meson $D_{s,2c}$.  
}
\label{tab:hsresults}
\begin{tabular}{llllll}
\hline \hline
Set & $aM^{\text{QCD}}_{D_{s,2c}}$ & $R_{D_{s,2c}}$ & $\delta \Delta_{\mathrm{FV}}$ [MeV] & $A_{D_{s,2c}}$ [MeV] \\
\hline
5$\ddagger$ & 1.34202(14) & 0.000637(43) &  0.3305(20) & 5.06(30) \\
6$\ddagger$ & 0.91076(19) & 0.000489(75) &  0.6682(44) & 4.84(51) \\
\hline \hline
\end{tabular}
\end{table}

\begin{figure}
\centering
\includegraphics[width=0.45\textwidth]{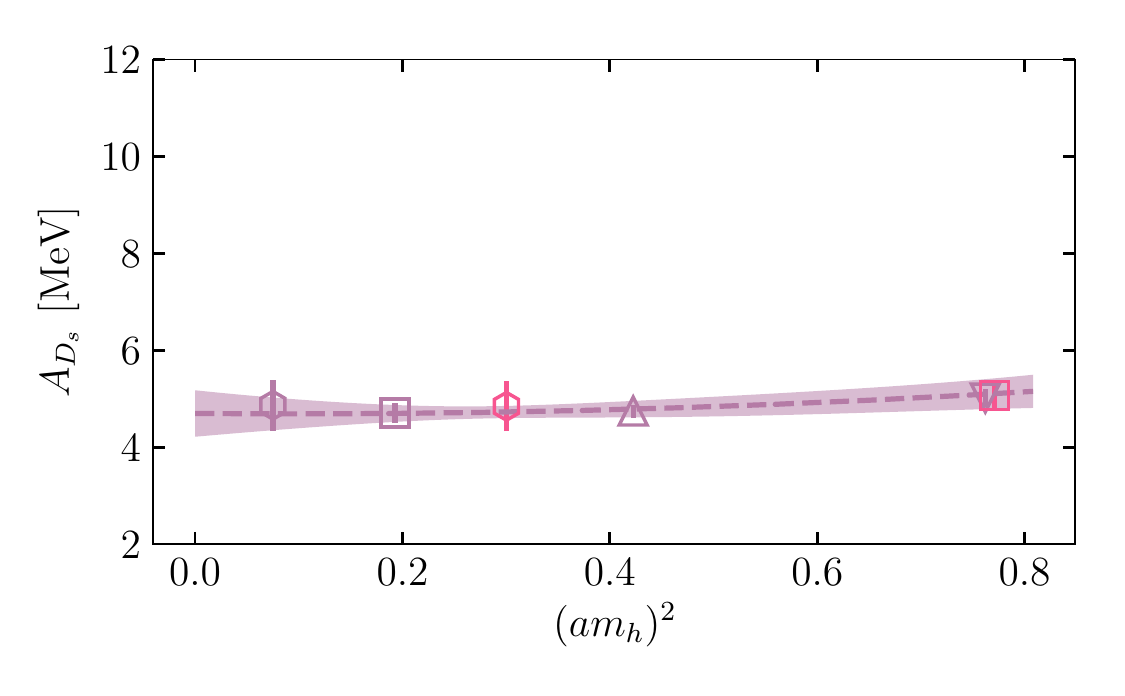}
\caption{The coefficient, $A$, of the QED interaction effect 
(Eq.~(\ref{eq:QED-mass-shift})) in the 
$D_s$ meson mass, shown with 
purple symbols as a function of squared lattice spacing (in units 
of the heavy quark mass, here $m_c$).  
The fit is described in the text. The pink symbols show the same 
results, but for mesons made from a heavy quark and strange antiquark with 
heavy quark 
mass twice that of the $c$ quark. Symbol shapes match those of the $D_s$ 
results on the same gluon field configurations. 
}
\label{fig:Ds}
\end{figure}

The results for the QED interaction coefficient $A$ for this case 
are shown in Fig.~\ref{fig:Ds}. 
$A_{D_s}$ is positive, and combined with a positive product of electric 
charges gives, as expected, a positive shift to the meson 
mass because the Coulomb interaction inside 
an electrically charged meson is repulsive. 
We perform the same continuum extrapolation fit as for the charmonium 
case, with the same priors, see Eq.~(\ref{eq:amc-fit}). 
Our fit returns a value 
\begin{equation}
\label{eq:dsresult}
A_{D_s}= 4.69(48) \, \mathrm{MeV}
\end{equation}
with a $\chi^2/\mathrm{dof}$ of 0.015. 

We can also, as in the heavyonium case, work with heavy quarks with 
mass $2m_c$. The results for this mass are given in Table~\ref{tab:hsresults} and also plotted 
in Fig.~\ref{fig:Ds}. 
In contrast to the heavyonium case, we find that the coefficient $A$ hardly 
changes as we change the heavy quark mass in the heavy-strange meson 
to be $2m_c$. 
From a fit to this case we obtain
\begin{equation}
\label{eq:2dsresult}
A_{D_{s,2c}}= 4.68(66) \, \mathrm{MeV}
\end{equation}
with $\chi^2/\mathrm{dof}$ of 0.002.
This agrees very well with that for the $D_s$ above, consistent with 
the fact that the points are on top of each other in Fig.~\ref{fig:Ds}. 

\section{Discussion}
\label{sec:discussion}

The coefficient $A$ is a physical quantity, encoding 
information about meson structure. 
The quantitative information that lattice QCD results 
for $A$ provide 
can be used to calibrate more qualitative model approaches 
for comparable quantities. 
We discuss this below first for heavyonium and then heavy-light mesons. 

The language of potential models provides a reasonably good 
approximation for heavyonium. 
A simple Cornell potential~\cite{Eichten:1979ms} of the form 
\begin{equation}
\label{eq:cornell}
V(r) = -\frac{\kappa}{r} + \frac{r}{b^2}
\end{equation}
can readily be tuned to give the radial excitation energy of charmonium 
with an accuracy of $\sim$10\%. Here $\kappa$ is $4\alpha_s/3$ and 
$b^2$ is the inverse string tension. The parameters used 
are: $\kappa=0.52$ and $b=2.34\,\mathrm{GeV}^{-1}$, along with a 
$c$ quark mass in the kinetic energy 
term of Schr\"{o}dinger's equation of 1.84 GeV~\cite{Eichten:1979ms}. 
It 
is then straightforward to perturb the 
coefficient of $1/r$ in Eq.~(\ref{eq:cornell}) by $\alpha_{\mathrm{QED}}$   
to include the Coulomb interaction effect and determine a value 
for the ground-state energy shift which is the potential model 
value for $A$ for charmonium, $A_c^{\mathrm{potl}}$. 
Alternatively this can be obtained 
by integrating over $\alpha_{\mathrm{QED}}/r$ weighted by the square 
of the ground-state wavefunction. 

Doing this gives a 
value for $A_c^{\mathrm{potl}}$ of 5.9 MeV 
(this is a shift of 2.6 MeV downwards in the meson mass 
when multiplied by $e_{q_1}e_{\overline{q}_2}$ for 
charmonium~\cite{Davies:2009tsa}). 
This result is for the leading spin-independent central potential
of Eq.~(\ref{eq:cornell}) and does not include any spin-dependent effects. 
Our lattice QCD results, on the other hand, are for 
the $\eta_c$ and $J/\psi$ mesons
separately. To compare our lattice results to those 
from a spin-independent potential 
we need to take the spin average: 
\begin{equation}
\label{eq:spinav}
A_c = \frac{A_{\eta_c}+3A_{J/\psi}}{4} 
\end{equation} 
Our results from Section~\ref{sec:lattice} (Eq.~(\ref{eq:charmres})) yield 
\begin{equation}
\label{eq:Ac}
A_c =  5.12(17) \, \mathrm{MeV}
\end{equation}

The potential model result, $A_c^{\mathrm{potl}}$, given above 
is 15(3)\% larger than our lattice QCD value. The uncertainty 
here comes from the lattice QCD calculation where it can be quantified. 
Clearly more sophisticated potential 
models, including potentials derived from lattice QCD~\cite{Bali:2000gf}, could be 
used to improve on 
the potential model result. Our value for $A_c$ in Eq.~(\ref{eq:Ac}) 
can also be used to tune the parameters of potential models. 
Frequently the tuning is done using 
quantities such as the wavefunction at the origin, 
along with the spectrum (see, for example,~\cite{Eichten:1995ch, Eichten:2019hbb}). 
The wavefunction at the origin is not a physical 
quantity, however, and there are sizeable uncertainties associated with 
renormalising this to relate it to experimental decay rates. 
In contrast the quantity $A$ is a physical, 
renormalisation-group invariant quantity that can be compared much more 
precisely. A systematic uncertainty of order 10\% on 
$A_c^{\mathrm{potl}}$ might be expected on 
the potential model result from missing $\mathcal{O}(v^4)$ relativistic 
corrections. However this could be ameliorated by tuning the potential. 

We can also compare the lattice and Cornell potential results for the 
heavier quark mass of $2m_c$. Then our lattice spin-averaged result, 
using the values from Eq.~(\ref{eq:2charmres}) is 
\begin{equation}
\label{eq:Ac2}
A_{2c} =  6.71(49) \,\mathrm{MeV} . 
\end{equation}
The result for $A_{2c}^{\text{potl}}$ from the same Cornell potential as 
for the $m_c$ case is 
9.1 MeV, now 30\% too large. 

Our results show a variation of $A$ with quark mass that behaves
approximately as $\sqrt{m}$. We can compare this to what might 
be expected from scaling arguments for a potential of the 
form $Cr^N$. Then, as we change the quark's 
reduced mass, $\mu \equiv m/2$, 
we obtain the same solution for a rescaled 
distance $\lambda r$ where~\cite{Quigg:1979vr, Davies:1997hv} 
\begin{equation}
\lambda \propto \mu^{-1/(2+N)} \, .
\end{equation} 
A $\sqrt{\mu}$ behaviour for 
$A^{\mathrm{potl}} \equiv \langle \alpha_{\mathrm{QED}}/r \rangle$ 
would then correspond to $N\approx 0$. 
Such a form for the heavy quark potential is in fact a standard one 
that 
has been successful in obtaining spectra, either taking $N$ to 
be a small value or taking $V(r)$ to be 
logarithmic~\cite{Martin:1980jx, Quigg:1977dd}. 
These forms for the potential give a wavefunction that does not 
grow so rapidly with mass at small distance as the Cornell potential 
and might give results for $A_c^{\mathrm{potl}}$ and $A_{2c}^{\mathrm{potl}}$ 
in better agreement 
with our lattice QCD value. 
See~\cite{Eichten:1994gt, Eichten:1995ch} for a comparison of spectrum 
and wavefunction results for 
different potential forms. 

The difference of our results for $A_{\eta_c}$ and $A_{J/\psi}$ gives 
the `direct' QED effect on the charmonium hyperfine splitting, when multiplied 
by -4/9. Note that in~\cite{Hatton:2020qhk} we also included a quark-line 
disconnected contribution from QED to the hyperfine splitting 
that is not included here. 
We have  
a difference of $A_{J/\psi}$ and $A_{\eta_c}$ of -2.5(3) MeV from 
Eq.~(\ref{eq:charmres}) for the $m_c$ case and -2.4(8) MeV from 
Eq.~(\ref{eq:2charmres}). The QED effect is then to raise the vector 
mass with respect to the pseudoscalar by about 1 MeV in both cases (using 
electric charge $2/3$). 

The hyperfine splitting itself falls with increasing quark mass, so 
the relative QED effect (for the same electric charge) is growing. 
In~\cite{Hatton:2020qhk} we did a complete analysis of the charmonium 
hyperfine splitting, including QED effects, as a function of lattice 
spacing and sea quark mass. To compare the $m_c$ and $2m_c$ cases here 
it is sufficient to 
take results from a single ensemble, set 6, as a guide to variation 
with heavy quark mass. 
The results in Tables~\ref{tab:charm-results} and~\ref{tab:heavy-results} 
then yield a pure QCD hyperfine splitting of 111 MeV for the $m_c$ 
case and 75 MeV 
for the $2m_c$ case, i.e. a fall of 30\% on doubling the quark mass. 
This is to be compared to a QED contribution that does not change 
(at the level of our uncertainties). 
A key difference between the QCD and QED hyperfine splittings 
is the effective inclusion 
(implicit in our lattice QCD calculation) of a running coupling constant 
in the QCD case which reduces the splitting as the mass increases.  

The QED hyperfine effect can also be compared to the expectation 
from a potential model calculation, by determining the impact of the 
perturbation from the Coulomb term on the wavefunction at the origin.    
The leading term in the hyperfine splitting from a potential model 
is proportional to the square of the wavefunction at the origin and so 
the percentage change in the hyperfine splitting is simply twice the 
percentage shift in $\psi(0)$. 
For the Cornell potential discussed above we find the percentage 
change in the hyperfine splitting (for $e_{q}=1$) to be 1.92\% for the $m_c$ case 
and 2.74\% for the $2m_c$ case. This shows an increase in the percentage 
QED effect that grows with the quark mass, as we find from our lattice 
calculation. 
To compare more quantitatively to our results 
we multiply these percentages by the pure QCD 
hyperfine splitting on set 6 given above. 
This gives a QED hyperfine effect (for $e_{q}=1$) 
of 2.1 MeV for both cases, in good agreement with our results 
from the difference of $A_{\eta_c}$ and $A_{J/\psi}$. 
Note, however, that sizeable ($\mathcal{O}(30\%)$ for charmonium) 
systematic errors are to be expected 
in analyses of fine structure from a potential model, 
so a semi-quantitative 
comparison is the best we can do here. 

We now turn to the heavy-light meson case. 
In~\cite{Davies:2010ip} we analysed a model of QED and light-quark 
mass effects in heavy-light pseudoscalar meson masses to isolate the QED
interaction term phenomenologically.  We used~\cite{Goity:2007fu} 
\begin{equation}
\label{eq:hl-pheno}
M(e_{q_h},e_{\overline{q}_l},m_q) = M_0 + Ae_{q_h}e_{\overline{q}_l} + Be_{q_l}^2 + C(m_{q_l}-m_l) 
\end{equation}
where $e_{q_h}$ and $e_{\overline{q}_l}$ are the electric charges of the heavy quark  
and light antiquark respectively and $m_{q_l}$ is the light quark mass, 
$m_l$ being the average $u/d$ quark mass. The coefficient $A$ gives 
the QED interaction term that we are interested in here. The 
coefficient $B$ is that of the light-quark QED self-energy, assumed to be 
independent of light quark mass. No term is included for the heavy 
quark self-energy because that cancels in the differences of heavy-light
meson masses for the same heavy quark that we will use to fix the 
coefficients. The coefficient $C$ allows for linear dependence on 
the light quark mass, independent of QED effects. 
From heavy quark symmetry we can expect that 
$A$, $B$ and $C$ will be constant up to $\Lambda/m_h$ corrections as 
the heavy quark mass
$m_h \rightarrow \infty$ and independent of $m_{q_l}$ up to small 
chiral corrections. This model was also used in~\cite{Bazavov:2017lyh}.  

If we assume that the coefficient $A\equiv A^{\mathrm{phen}}$ is independent of heavy quark 
mass (i.e. we ignore $\Lambda/m_h$ terms) we can easily determine 
it from experimental 
information. If we add the experimental
mass difference of $B^+$ and $B^0$ to the mass 
difference of $D^+$ and $D^0$~\cite{Zyla:2020zbs} then the terms with coefficients 
$C$ and $B$ (if independent of heavy quark mass) cancel out. We have 
\begin{eqnarray}
\label{eq:Aexp}
\frac{2}{3}A^{\mathrm{phen}} + \frac{1}{3}A^{\mathrm{phen}} &=& 4.822(15) - 0.31(5) \, \mathrm{MeV} \, , \nonumber \\
A^{\mathrm{phen}} &=& 4.51(5) \, \mathrm{MeV} . 
\end{eqnarray}
This result agrees well with our lattice determination of 
$A$ for the $D_s$ meson in Eq.~(\ref{eq:dsresult}). 

From our calculation of results with a heavy quark mass twice that 
of charm (Eq.~(\ref{eq:2dsresult})) we are able to show that 
indeed $A$ is independent of heavy 
quark mass at the level of our uncertainties ($\sim$10\%). 
It would be straightforward to extend our calculation to the 
$D$ meson from the $D_s$ to test for any dependence 
of $A$ on the light quark mass. 

\section{Conclusions}
\label{sec:conclusions}

We have shown here how to separate out the QED interaction piece from 
the self-energy terms in the determination of the effect of including QED 
for valence quarks on heavyonium and heavy-light meson masses. 
Lattice QCD calculations are now accurate enough that the effect of 
QED, at least for the valence quarks, can have an impact. The full 
effect of QED needs to be included in order to tune parameters, such 
as quark masses, by tuning meson masses until they take their experimental 
value in the QCD+QED calculation (this was done, for example, for 
QCD + quenched QED in~\cite{Hatton:2020qhk}). 

There are multiple reasons for wanting to separate out the 
QED interaction piece from the self-energy terms of the QED effect, however. 
One is to 
test our understanding of the physical contribution of QED by comparing 
to phenomenological model calculations. 
Another is to use the effect 
as a probe of meson, and more generally, hadron structure by using it 
to determine an effective average radial separation of the valence 
quarks. 

We have determined the coefficient $A$ of the QED interaction 
piece 
for the $\eta_c$, $J/\psi$ and $D_s$ mesons as well as for the corresponding 
mesons constructed by doubling the $c$ quark mass 
(see Eqs~(\ref{eq:charmres}),~(\ref{eq:2charmres}),~(\ref{eq:dsresult}) 
and~(\ref{eq:2dsresult})). 
The uncertainties we obtain at the physical point are 5\% for the heavyonium 
case and 10\% for heavy-light. 

A simple potential model gives results for the Coulomb interaction 
effect in charmonium 
in reasonable agreement with the lattice QCD numbers 
(spin-averaged to remove 
spin effects). We suggest 
that the lattice results
could be used to tune potential models more accurately. This in turn 
could improve results for calculations, for example involving excited 
states and hadronic decay channels, that are currently more readily done in a 
potential model than using lattice QCD. We also find that a phenomenological 
model based on heavy quark symmetry gives good agreement 
with our $D_s$ results. We
are able to demonstrate in that case that $A$ is independent of heavy 
quark mass.    

Since $A$ is dominated by the Coulomb interaction effect 
for heavy mesons we can define an effective size parameter 
$\langle 1/r_{\mathrm{eff}} \rangle$ by dividing 
our results for $A$ by $\alpha_{\mathrm{QED}}$. 
This gives values for $\eta_c$, $J/\psi$ and $D_s$ 
mesons of 
\begin{eqnarray}
\label{eq:size}
1/\langle 1/r_{\mathrm{eff}} \rangle &=& 0.206(8) \, \mathrm{fm}, \quad \eta_c \\
&=& 0.321(14) \, \mathrm{fm}, \quad J/\psi \nonumber \\
&=& 0.307(31) \, \mathrm{fm}, \quad D_s \, .\nonumber 
\end{eqnarray}
The $\eta_c$ result can be compared to the 
value for $\sqrt{\langle r^2 \rangle}$ using 
$\langle r^2 \rangle = 6/M_{J/\psi}^2$ which is in reasonable agreement 
with our results for the electromagnetic form factor of the
$\eta_c$ at small squared momentum-transfer, $q^2$~\cite{Davies:2019nut}. 
This would give $\sqrt{\langle r^2 \rangle}$ = 0.156 fm. 
We also find that the size parameter from Eq.~(\ref{eq:size}) 
falls for heavier heavyonium 
masses approximately as $\sqrt{m}$ but does not change at all 
for heavy-light mesons as the mass is increased. 

We believe that this could be a useful approach to assessing 
the size of other hadrons because it requires only the 
calculation and fitting of correlated 2-point functions. 
The noncompact QED action is simply being used as a convenient 
way to probe the $r$-dependence so
a larger value of $\alpha_{\mathrm{QED}}$ than the physical one 
can be used to increase the signal for the perturbation~\cite{Duncan:1996xy}. 
For these purposes it might also be easier to use a purely Coulomb photon 
on each timeslice of the lattice as the direct Fourier transfrom of $1/r$. 
By giving electric charge 
to pairs of quarks in more complicated hadrons such as baryons, tetraquarks 
or pentaquarks it might be possible 
to distinguish diquark-like configurations where they occur. 
We plan to test this out. 

\subsection*{\bf{Acknowledgements}}

We are grateful to the MILC collaboration for the use of
their configurations and QCD code. We adapted this to include quenched QED.  
Computing was done on the Cambridge service for Data 
Driven Discovery (CSD3), part of which is operated by the 
University of Cambridge Research Computing on behalf of 
the DIRAC 
HPC Facility of the Science and Technology Facilities 
Council (STFC). The DIRAC component of CSD3 was funded by 
BEIS capital funding via STFC capital grants ST/P002307/1 and 
ST/R002452/1 and STFC operations grant ST/R00689X/1. 
DiRAC is part of the national e-infrastructure.  
We are grateful to the CSD3 support staff for assistance.
Funding for this work came from the UK
Science and Technology Facilities Council grants 
ST/L000466/1 and ST/P000746/1 and from the National Science 
Foundation.

\bibliography{coulomb}

\end{document}